\documentclass{revtex4}

\usepackage{epsfig}
\input axodraw.sty

\newcommand{\mett}{{\not\!\!E}_{T}}

\def\beq{\begin{equation}}
\def\eeq{\end{equation}}
\def\bea{\begin{eqnarray}}
\def\eea{\end{eqnarray}}
\def\sriimge{0.098}
\def\sriige{0.36}
\def\nriimge{195}
\def\nriige{720}

\begin{document}
\bibliographystyle{apsrev}
\hfill\parbox{8cm}{\raggedleft CERN-TH/2001-303 \\ hep-ph/0111014 \\
}

\title{Explaining Anomalous CDF \boldmath{$\mu \gamma$} Missing-\boldmath{$E_T$} Events With Supersymmetry}
\author{B.C.  Allanach, S. Lola and K. Sridhar}

\altaffiliation{On leave of absence 
from the Tata Institute of Fundamental Research, Homi Bhabha Road, 
Mumbai 400 005, India.}
\affiliation{CERN, Geneva 23, CH 1211, Switzerland}
\date{\today}

\begin{abstract}
CDF recently reported an excess of events in the $\mu \gamma$
missing $E_T$ ($\mett$) channel that disagrees with the Standard Model prediction.
No such excess was observed in the $e
\gamma \mett$ channel. We explain the excess via resonant smuon production 
with a single dominant R-parity violating coupling $\lambda'_{211}$,
in the context of models where
the gravitino is the lightest supersymmetric particle.
The slepton decays to the lightest neutralino
and a muon followed by neutralino decaying to a gravitino and photon.
We determine a viable region of parameter space that fits the kinematical
distributions of the Run I excess
and illustrate the effect by examining the best fit point in detail.
We provide predictions for an excess in the $\mett$
and photon channel at Run I and Run II. Run II will decisively rule out or
confirm our scenario.
\end{abstract}

\maketitle

\section{Introduction}

CDF has recently presented results 
on the production of combinations
involving at least one photon and one lepton ($e$ or $\mu$)
in $p{\bar p}$ collisions at $\sqrt{s}= 1.8 ~{\rm TeV}$,
using 86 pb$^{-1}$ of Tevatron 1994-95  data \cite{CDF}.
In general the  results were consistent with the Standard Model (SM),
however 16 photon-lepton events with
large ${{\not\!\!E}_{T}}$ were observed, with
$7.6\pm0.7$ expected.
Moreover, 11 of these events involved muons
(with 4.2 $\pm$ 0.5 expected)
and only 5 electrons (with 3.4 $\pm$ 0.3 expected), 
suggestive of a lepton flavour violating asymmetry involving muons.

What can such a process be?
A natural framework with explicit flavour violating couplings is provided
by R-violating supersymmetry \cite{pheno},
which contains operators 
with a complicated flavour structure in the superpotential 
\beq
W_{RPV}=\frac{1}{2}\lambda_{ijk} L_iL_j{\bar E}_k+\lambda'_{ijk}L_iQ_j{\bar D_k}+
\frac{1}{2}\lambda''_{ijk}{\bar U_i}{\bar D_j}{\bar D_k}+\mu_i L_i H_2
\label{eq:superpot}
\eeq
where $L$ $(Q)$ are the left-handed lepton (quark) superfields while ${\bar
E},{\bar D},$ and ${\bar U}$ contain the corresponding right-handed fields,
and $i,j,k$ generation indices.
The second of the above terms is of particular interest, since it can
lead to resonant slepton production in hadron-hadron collisions 
\cite{previous}, via the diagram that appears below.
\begin{figure}
\begin{center}
\begin{picture}(200,70)
\ArrowLine(30,5)(60,25)
\ArrowLine(60,25)(30,45)
\DashLine(60,25)(90,25){3}
\ArrowLine(90,25)(120,5)
\DashLine(90,25)(120,45){3}
\DashLine(120,45)(150,35){3}
\Photon(120,45)(150,65){3}{2}
\put(32.5,2.5){$q$}
\put(32.5,45){$\bar q'$}
\put(122.5,5){$\mu$}
\put(100,40){$\chi_0$}
\put(75,32.5){$\tilde\mu$}
\put(152.5,32.5){$\tilde G$}
\put(152.5,62.5){$\gamma$}
\end{picture}
\end{center}
\caption{Resonant smuon production and subsequent decay}
\label{feynman}
\end{figure}
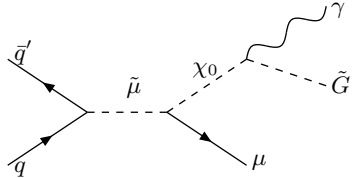

Such a resonance would lead to enhanced cross sections with a rich final state
topology, which, as we are going to show, 
can explain the CDF anomaly. What would then be the structure
of the associated operator? R-violating couplings have upper bounds coming from
various flavour-violating processes \cite{constraints}. Therefore, to get
the requisite number of events to explain the observed anomaly, a sizable
cross section is required which would then imply a process with valence quarks 
in the initial state. Since the events
are seen in the muon channel, the operator can be specified to be
$L_2Q_1\bar{D}_1$, which generates the couplings ${\tilde \mu}u\bar{d}$ and
${\tilde \nu}_\mu d \bar{d}$ (and charge conjugates), along with other
supersymmetrised copies involving squarks. 
This coupling, $\lambda'_{211}$, is constrained from  
$R_\pi = \Gamma (\pi \rightarrow e \nu) / (\pi \rightarrow \mu \nu)$
\cite{bgh} to be 
$< 0.059 \times \frac{m_{\tilde{d_R}}}{100 ~{\rm GeV}}$ \cite{constraints}.

Upon production, sleptons 
(in our case, smuons or sneutrinos) can in general decay via a large
variety of channels \cite{previous} if they are kinematically accessible. 
However, the crucial observation is that R-violating supersymmetry by itself
may not account for the observed anomaly, because of the fact that the
anomaly is observed in a channel where a photon is produced.
However, if the gravitino (present in all models where supersymmetry is
gauged) is the lightest supersymmetric particle (LSP)
it is too long-lived to decay within the detector \cite{gaugmed}.
Thus, the gravitino, $\tilde G$, provides the missing energy signature since it is
electrically neutral and interacts rather weakly with matter.
If the neutralino, as is often the case, is dominantly
photino, then the decay $\chi_1^0 \rightarrow {\tilde G} \gamma$ can dominate
\cite{comm}. It is interesting to note that the
$ee \gamma \gamma \mett$ event
recorded by CDF~\cite{abe} can be explained
by such a decay~\cite{kane}.

Since at the moment there is neither enhancement in the two-fermion
final state, nor observation of chains of cascade decays, the most
natural explanation is that the R-conserving
decay mode of the smuon which produces the lightest neutralino dominates over
the rest while subsequently
$\chi^0_1 \rightarrow {\tilde G} \gamma$. The competing R-parity violating
decay modes of $\chi_1^0 \rightarrow \nu jj$ and $\chi_1^0 \rightarrow \mu jj$ 
leading to $\mu jj \mett$ or $\mu \mu jj$ final states become negligible (as
is the case here) when $\lambda'_{211}$ and $m_{\tilde G}$ are both small
enough. 
Smuon decay into two jets via the R-parity violating mode 
is essentially unobservable 
because of the huge 2 jet background. For example, for a resonance mass of
200 GeV, only a $\sigma . B> 1.3 \times 10^4$ pb is excluded at 95\%
C.L.~\cite{cdfjets}. 
This will not provide a restrictive bound upon our scenario.

It is worth stressing the clarity of the signatures, but also of future
predictions in the case of a resonant process. 
Moreover, the presence of both slepton and sneutrino resonances are in
principle to be expected, and we provide a prediction for 
$\gamma {{\not\!\!E}_{T}}$ events.
The higher statistics in Run II of the Tevatron should allow verification
our model.

The new aspect of the model we present here compared to previous studies
of resonant slepton production at hadron colliders~\cite{previous},
is to marry the gravitino LSP scenario with R-parity violating supersymmetry. 
This marriage has been considered before in the context of dark
matter~\cite{contexts}.

\section{Model and Results}

We use the {\small \tt ISASUSY} part of the {\small \tt ISAJET7.58}
package~\cite{isajet} to generate the spectrum, branching ratios and 
decays of the sparticles. 
For an example of parameters, we choose (in the notation used by
ref.~\cite{isajet}) $\lambda'_{211}=0.01$, $m_{3/2}=10^{-3}$ eV, $\tan
\beta=10$, $A_{t,\tau,b}=0$, and scan over the bino mass
$M_1$ and the slepton mass
$m_{\tilde l}\equiv m_{{\tilde L}_{1,2}}=m_{{\tilde e}_{1,2}}$
GeV. The values of $\lambda'_{211}$ and $m_{3/2}$ are dictated by the need
to have the decays shown in Fig.~\ref{feynman} being dominant. However, there
are ranges of values in the $R$-violating coupling and the gravitino
mass where this decay chain is obtained. In fact, the acceptable ranges are
an order of magnitude
in $\lambda'_{211}$ and two orders of magnitude
in $m_{3/2}$. $\mu$ together with 
other flavour diagonal soft supersymmetry breaking parameters are set to
be so heavy that any superparticles except the first two
generation 
sleptons, the lightest neutralino and the gravitino are too heavy to be
produced or to contribute to cascade 
decays in Tevatron data. They therefore do not appear in this
analysis. We have checked that this is true over a large volume of parameter
space. 
We emphasise that this is a representative hyperplane in the
supersymmetric parameter space and not a special choice. 

We use {\small \tt HERWIG6.3}~\cite{herwig} including parton showering (but
not including isolation cuts) to calculate cross-sections for single slepton
production. 
A $\gamma$-in-active-region cut requires that the photon {\em
not} 
have rapidity $|\eta|>1$ or $|\eta|<0.05$.
The region $0.77<\eta<1.0, 75^\circ<\phi<90^\circ$ is also excluded because it
is not instrumented. Fiducial photon detection efficiency was set to be
81$\%$, whereas for the muons it is 66$\%$ for $1.0>|\eta_\mu|>0.6$ and 45$\%$
for $\vert \eta \vert < 0.6$. $\mett$ and the $E_T$ of both the muon and 
photon were required to be greater than 25 GeV. 

\begin{figure}
\unitlength=1in
\begin{picture}(6,2.3)
%\put(1.35,0)%{\includegraphics{etmiss.eps}}
%{\epsfig{file=muon.eps, width=2.7in}}
\put(0,0)%{\includegraphics{muon.eps}}
{\epsfig{file=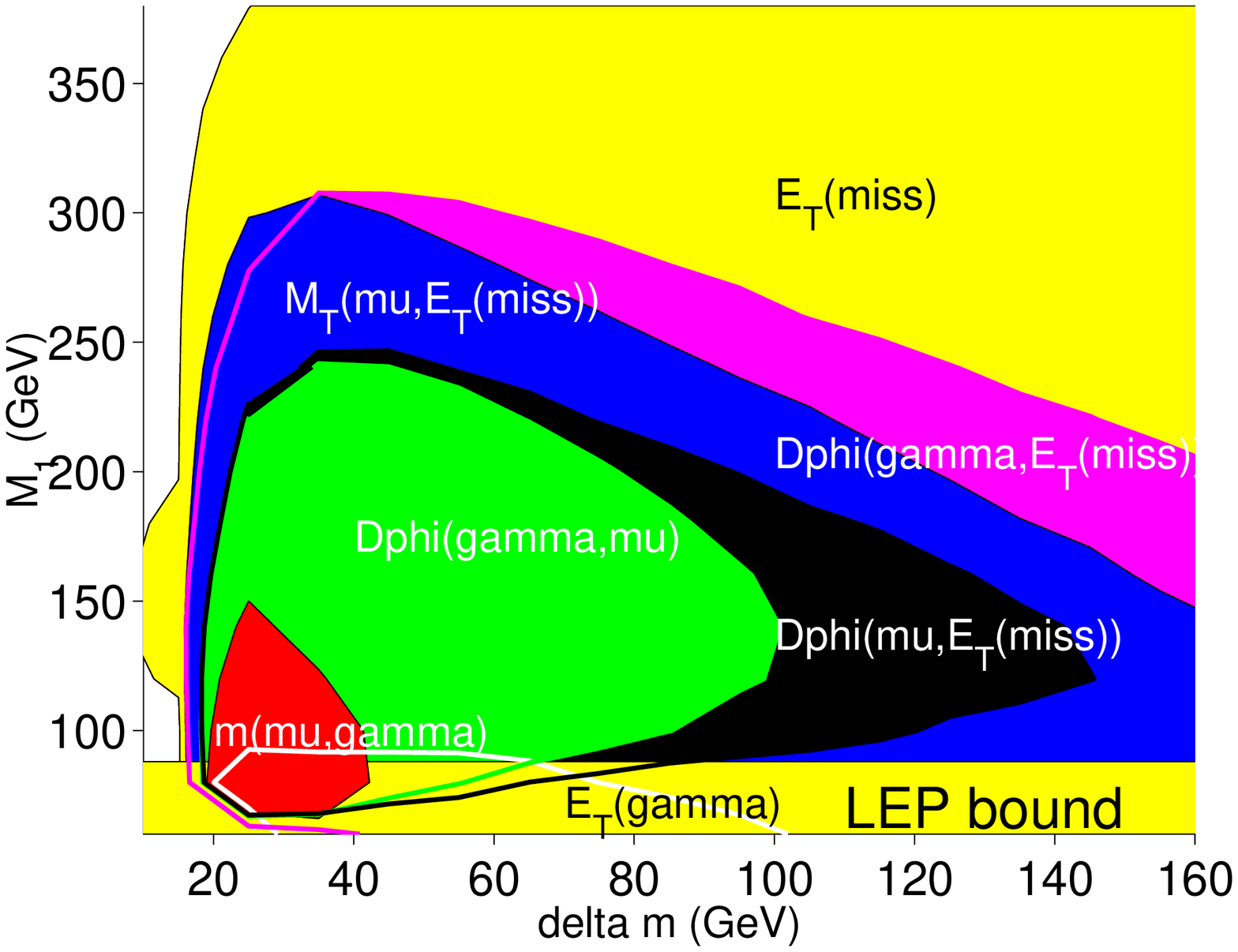, width=3in}}
\put(3,0)%{\includegraphics{photon.eps}}
{\epsfig{file=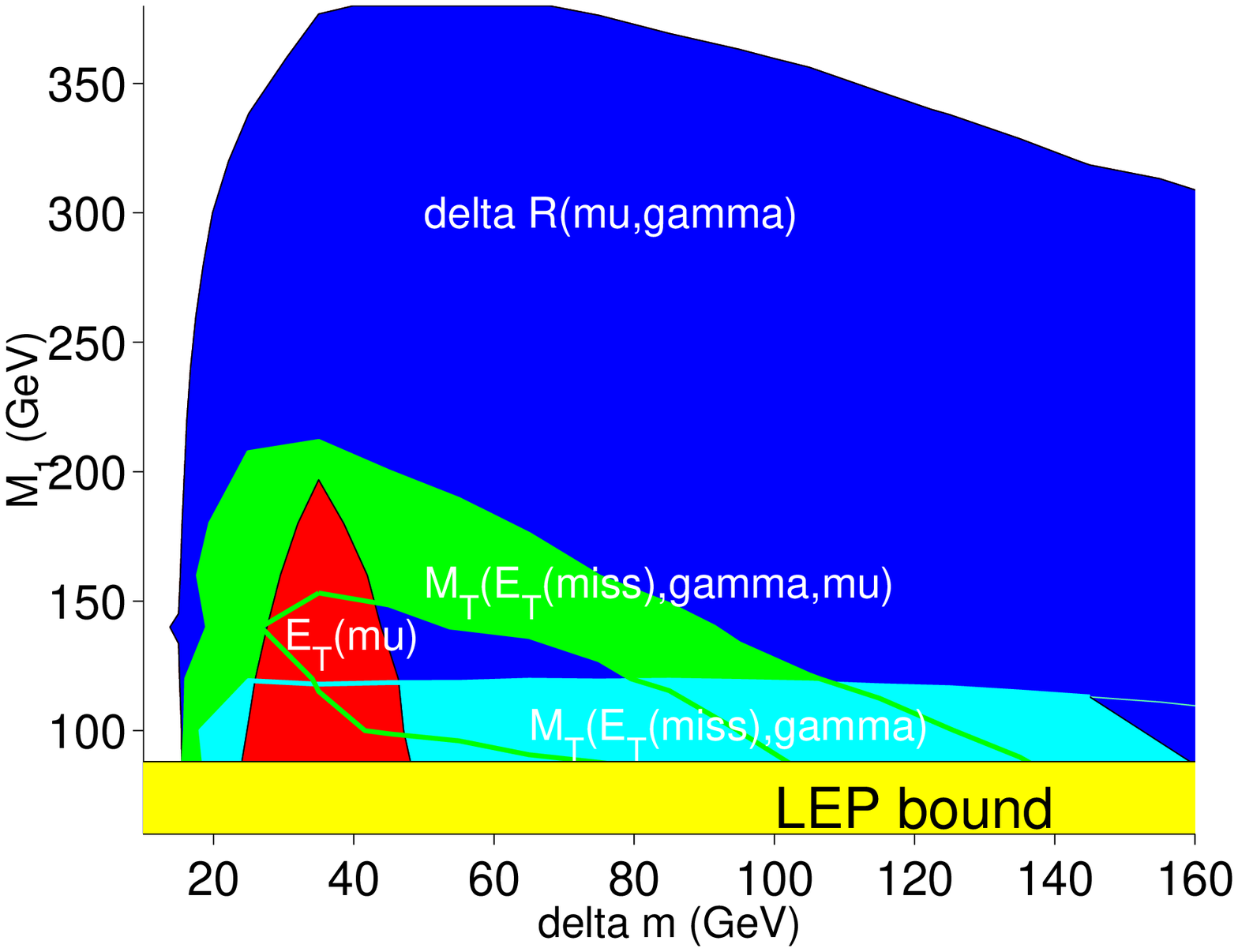, width=3in}}
\end{picture}
\caption{Scans over $M_1$ and $\Delta m$. The 95$\%$
C.L. regions indicated by the fit to each kinematical distribution is shown.}
\label{scans}
\end{figure}
We calculate the difference in log likelihood  between our model and the
SM given by each kinematical
variable that was presented in ref.~\cite{CDF}. This provides 95$\%$
C.L. limits upon $M_1$ and $\Delta m=m_{\tilde l}-M_1$. We show the viable
regions for the energy distributions: $E_T(\gamma,\mu)$, $\mett$, the mass $m$
and transverse mass $M_T$ 
distributions 
$m_{\mu \gamma}$, $M_T(\mu \mett)$, $M_T(\mett \gamma)$, $M_T(\gamma \mu
\mett)$ and various transverse angular separations $\Delta\phi_{ij}$, where
$i,j=\mu, \gamma, \mett$. $\Delta R$, defined as the distance in $\eta-\phi$
space between the muon and the photon, is also used. 
It is not possible to take correlations between these different kinematical
variables into account because we do not possess the multi-dimensional data.
Therefore we resort to examining each one in turn and see to what extent each
region overlaps. Fig.~\ref{scans} shows that all of the 95$\%$ confidence level
regions overlap at $M_1\approx 90$ GeV, $\Delta m=25-40$ GeV, indicating that
our model is in good agreement with all of the observed kinematical properties
of the events. The region at the bottom the plots is ruled out by LEP2 from
neutralino pair production~\cite{pdg}.

The most discriminating kinematical variable is $E_T(\mu)$, which favors our
model over the SM at the 3.3$\sigma$ level at the best fit point
$M_1=87$ GeV and $\Delta m=35$ GeV. 
We refer to this point as ``the best fit
point'' from now on, and examine its properties more closely.
\begin{figure}
\unitlength=1in
\epsfig{file=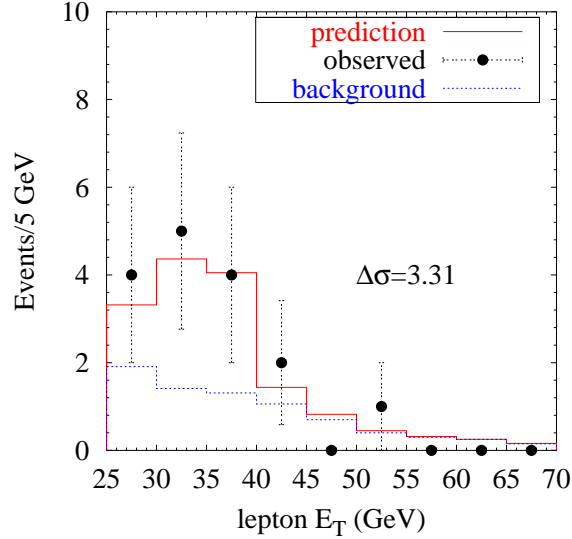, width=3in}
\caption{Lepton $E_T$ distribution at the best-fit point in the data (points),
SM 
background (dashed histogram) and our best-fit point (solid histogram).
$\sqrt{N}$ uncertainties have been imposed upon the data.}
\label{dists}
\end{figure}
We show the predicted distribution
of lepton $E_T$ in 
Fig.~\ref{dists} and compare it with the excess of the data over the
SM background. 
\begin{table}
\caption{MSSM spectrum used to explain anomalous events at the best fit
point within the acceptable fit range for $m_{\tilde G}=10^{-3}$ eV, $\tan
\beta=10$ 
and $\lambda'_{211}=0.01$. We have displayed the relevant 
sparticle masses. All other
sparticles are heavier than 1900 GeV, and thus do not interfere with our
analysis.}
\label{tab:spec}
\begin{tabular}{c|cccc} 
particle & ${\tilde e}_L$,${\tilde \mu}_L$ & 
${{\tilde \nu}_e}$,${{\tilde \nu}_\mu}$ & $\chi_1^0$ & ${\tilde
\mu}_R$,${\tilde e}_R$\\ \hline
best-fit mass & 131 GeV & 104 GeV & 87 GeV& 130 GeV \\
range & 121--162 GeV & 92--141 GeV & 87--120 GeV & 120--161 GeV \\
\end{tabular}
\end{table}
Important features of the
 sparticle spectrum are displayed in Table~\ref{tab:spec}.
We also show the range of sparticle masses corresponding to the acceptable
 fit range of 
 parameter space. The acceptable fit range is defined as being compatible with
 at least  all but one of the 95$\%$ C.L. regions in fig~\ref{scans}.
The relevant branching ratios of the smuon are
\begin{equation}
BR({\tilde \mu}_L \rightarrow \chi_1^0 \mu) = 0.984, \qquad
BR({\tilde \mu}_L \rightarrow \bar{u} d) = 0.015, \qquad
BR({\tilde \mu}_L \rightarrow {\tilde \mu} {\tilde G}) = 0.001,
\end{equation}
with a lifetime of $1\times 10^{-22}$ sec,
whereas for the lightest neutralino we have
\begin{equation}
BR(\chi^0_1 \rightarrow {\tilde G} \gamma) = 0.975, \qquad
BR(\chi_1^0 \rightarrow {\tilde G} e^- e^+) = 0.020, \qquad
\end{equation}
with a lifetime of $1 \times 10^{-18}$ sec. At such small values of
$\lambda'_{211}$ and $m_{\tilde G}$, R-parity violating decays of the 
lightest neutralino are negligible.
In Table~\ref{tab:res1}, we show the percentage of events making it through
each of the cuts. 
The table shows that 11.4$\%$ of the smuons produced end up as detected
$\mu \gamma \mett$ events in CDF. 
\begin{table}
\caption{Percentage of SUSY events for the best fit point
that satisfy cumulative cuts for $\mu \gamma \mett$ 
events at CDF, Run I. Events that pass a cut in a given entry also pass
those cuts to the left.}
\label{tab:res1}
\begin{tabular}{c|c|c|c|c|c|c}
cut & 
$E_T(\gamma)>25$ GeV & isolated detected $\gamma$ & $E_T(\mu)>25$ GeV&
$|\eta_\mu|<1.0$ & detected $\mu$ & 
$\mett>25$ GeV \\ \hline
percentage & 80.8 & 48.6 & 35.2 & 22.8 & 13.3 & 10.9 \\
\end{tabular}
\end{table}
The corrected cross-section of 0.091 pb
corresponds to 7.8 events additional to the 4.2$\pm0.5$ predicted
by the SM
for 86 pb$^{-1}$ of
luminosity, adequately fitting the excess of events quoted by CDF at Run I.

We now determine the rate of single sneutrino production at Run I.
The process is: ${\tilde \nu}\rightarrow\nu \chi^0_1$ followed by
$\chi^0_1 \rightarrow {\tilde G} \gamma$. This would appear
to mimic $Z \gamma$ production, where $Z \rightarrow \nu {\bar \nu}$.
To compute the cross-section for this process, we use the cuts used by
the D0 experiment in their $\gamma \mett$ analysis \cite {dzero}.
With their cuts, we predict a supersymmetric cross-section of $0.054$ pb for
the $\mett \gamma$ process at the Run I energy, which corresponds to about 0.7
events for the 14 pb$^{-1}$ data analyzed by the D0 experiment. The D0
experiment observed 4 events over a SM background of 1.8$\pm$0.2 events
but with a much bigger background coming from cosmic ray sources which
is estimated to be 5.8$\pm$1.0.
As far as we are aware, the analysis has not yet been done with the
full Run I Tevatron data but we would expect about 5.4 events for a
100 pb$^{-1}$ data sample. 

We perform the above analyses for Run II (at $\sqrt{s}=2$ TeV) for the best
fit point
in order to make predictions for 
observable supersymmetric cross sections:
\begin{equation}
\sigma(\gamma \mu \mett) = \sriimge \mbox{~pb},\qquad
\sigma(\gamma \mett) = \sriige \mbox{~pb},
\end{equation}
which, ought to be observable with good
statistics. Since the numbers for the cuts and the efficiencies
at Run II are not available, we have simply used those that the CDF experiment
used in 
their $\mu \gamma \mett$ analysis at Run I. To that extent, these numbers
are only indicative.
For example, 
with an integrated luminosity of 2 fb${}^{-1}$, these cross-sections would correspond
to \nriimge~and \nriige~events, respectively.
We predict 0.8 expected selectron pairs at Run I. 
Thus, the discrepancy with respect to the SM~\cite{abe} from the
observation 
of an $e e \gamma \gamma \mett$ event in the Run I data is vastly ameliorated.
R-parity conserving production processes such as these will be observable at
Run II providing more independent checks upon our scenario. 
One expects an identical number of smuon pairs, leading to a $\mu \mu \gamma 
\gamma \mett$ final-state. This final state has not yet been observed by CDF, 
but we note that combining the $e e \gamma \gamma \mett$ and $\mu \mu \gamma
\gamma \mett$ channels, our model still vastly ameliorates the discrepancy
with respect to the SM.

\section{Conclusions}

We have demonstrated that R-parity violating supersymmetry with a 
light gravitino can explain an anomalously high measured cross-section for
the $\mu \gamma \mett$ channel. We have provided possible tests for this
hypothesis, in the form of SUSY cross-sections for the 
$\gamma \mett$ channel, and predictions for the cross-sections of both
channels at Run II of the Tevatron collider. 
The $\gamma \mett$ channel looks particularly promising because it will allow
an 
independent check of our scenario. 

Another interesting question to ask is whether
the signal can also be obtained from a specific model of supersymmetry
breaking consistent with all other data.

\begin{acknowledgments}
We would like to thank K. Odagiri for focusing our attention on the $E_T$
distributions. 
\end{acknowledgments}


\begin{thebibliography}{99}
\bibitem{comm}
The radiative decay
$ \chi^{0} \rightarrow \nu \gamma$ is possible, but in practice
it is loop-suppressed with respect to the
3-body neutralino decays which leads to a different final
state.
\end{thebibliography}
\end{document}